# Modular Golden Gate Assembly of Linear DNA Templates for Cell-free Prototyping

**François-Xavier Lehr[1,2#] & Aukse Gaizauskaite[1,2,3#], Katarzyna Elżbieta Lipińska[1,2], Sara Gilles[1,2], Arpita Sahoo[1,2], René Inckemann[1], Henrike Niederholtmeyer[1,2,3*]**

[1]Max Planck Institute for Terrestrial Microbiology, Marburg, Germany

[2]Center for Synthetic Microbiology (SYNMIKRO), Philipps-Universität Marburg, Marburg, Germany

[3]Synthetic Biology, TUM Campus Straubing for Biotechnology and Sustainability, Technical University of Munich, Straubing, Germany

[#]FXL and AG contributed equally to this work.

*Correspondence to:

Henrike Niederholtmeyer: henrike.niederholtmeyer@tum.de

## Abstract

Cell-free transcription and translation (TXTL) systems have emerged as a powerful tool for testing genetic regulatory elements and circuits. Cell-free prototyping can dramatically accelerate the design-build-test cycle of new functions in synthetic biology, in particular when linear DNA templates are used. Here we describe a Golden Gate assisted workflow to rapidly produce linear DNA templates for TXTL reactions by assembling transcriptional units from basic genetic parts of a modular cloning toolbox. Functional DNA templates composed of multiple parts such as promoter, ribosomal binding site (RBS), coding sequence, and terminator are produced *in vitro* in a one-pot Golden Gate assembly reaction followed by PCR amplification. We demonstrate assembly and cell-free testing of promoter and RBS combinations, as well as characterization of a repressor–promoter pair. By eliminating lengthy transformation and cloning steps in cells and by taking advantage of modular cloning toolboxes, our cell-free prototyping workflow can produce data for large numbers of new constructs within a single day.

## 1. Introduction

In cell-free expression, the molecular machinery for transcription and translation is isolated from cells and supplemented with substrates as well as DNA templates to drive the synthesis of mRNA and proteins. Cell-free transcription and translation (TXTL) systems enable rapid prototyping of new genetic elements, such as promoters and ribosomal binding sites, as well as the engineering of transcription factors and riboswitches. TXTL reactions are typically performed in small volumes of a few microliters and fluorescent reporter outputs are measured in plate readers. This workflow allows the screening of libraries of genetic parts in high throughput *(1)*. Conveniently, it is possible to use linear DNA as expression templates to eliminate time-consuming cloning and transformation steps. To protect linear DNA from degradation by the RecBCD nuclease complex in *Escherichia coli* lysates, two approaches are commonly used. The first involves adding the phage protein GamS, which stabilizes linear DNA *(2)*. The second method includes short linear DNA strands containing repeats of the chi sequence, which act as competitive inhibitors, stalling the RecBCD nuclease *(3)*.

Libraries of linear DNA templates to screen can be assembled by different methods. One option is to synthesize them by PCR using overlapping oligonucleotides and PCR products *(4, 5)*. However, this approach may require the purchase of long oligonucleotide libraries and may encounter issues with non-specific PCR byproducts. Another option is to assemble expression templates from multiple basic parts by Golden Gate Assembly, followed by a PCR to amplify the transcriptional unit (TU) of interest *(2)*.

Golden gate cloning relies on Type IIS restriction enzymes that cleave outside their recognition sequence, generating short single stranded overhangs that can be customized to assemble fragments in a defined order *(6)*.The Modular Cloning (MoClo) system *(7)* uses standardized fusion sites that allow the assembly of TUs or multigene constructs from multiple basic genetic parts like promoters, ribosomal binding sites (RBS), coding sequences (CDS) and terminators in the correct order in a single reaction. Several toolboxes, such as the Marburg collection *(8)*, have been developed and contain collections of basic parts that are a useful starting point for many synthetic biology projects. In the Golden Gate cloning systems, basic parts like promoters, RBSs, and CDSs, are called level 0 parts and are stored on level 0 plasmids. These are then assembled into TUs, so called level 1 constructs. Golden Gate reactions are classically transformed into bacteria, where the assembled plasmid is propagated. However, an assembled level 1 TU can also simply be amplified by a PCR reaction with standard primers to produce linear DNA templates suitable for TXTL reactions. When starting with a collection of level 0 plasmids containing basic parts, large libraries of new constructs can be rapidly generated and characterized in a TXTL system.

The cell-free prototyping workflow we describe here (Fig. 1) allows TU assembly and measurements to be completed in a single day: 2 h Golden Gate reaction, 3 h PCR and DNA purification, and 5 h TXTL plate reader reaction. We also describe the assembly of new level 0 plasmids because it is often necessary to expand existing collections of level 0 parts with custom-designed parts for specific projects. Finally, to facilitate the screening of larger libraries, we report the setup of reactions in multiwell plates assisted by an automatic robotic liquid handling system capable of dispensing low volumes.

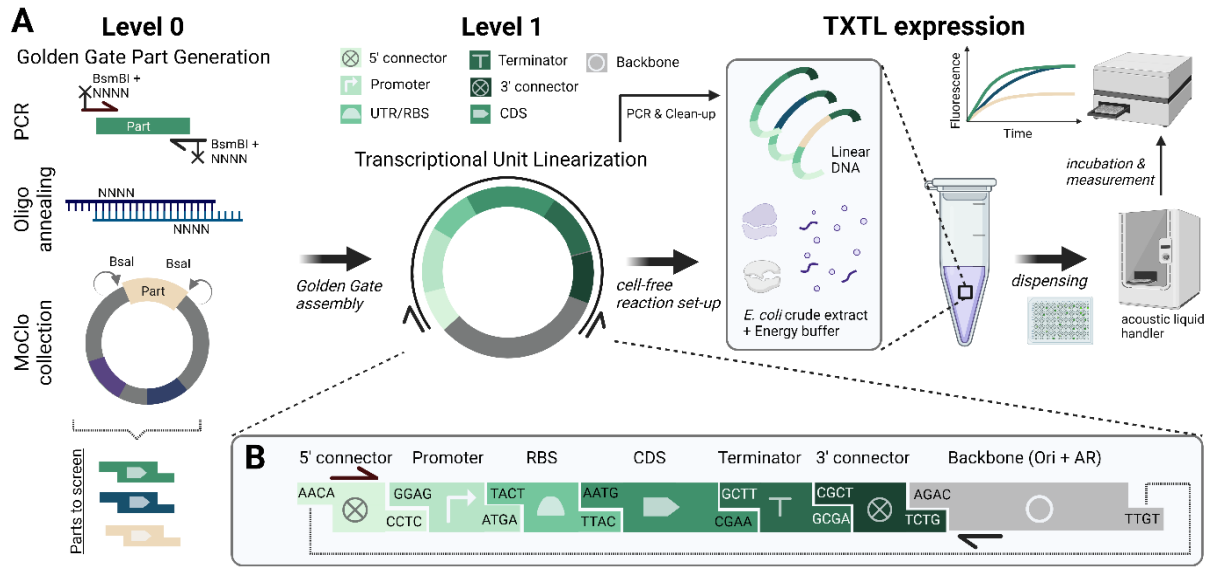

**Fig. 1** Overview of the cell-free prototyping workflow. A) Level 0 parts are assembled into level 1 constructs by Golden Gate Assembly and subsequently amplified and linearized by PCR for functional testing in cell-free reactions. B) Overview of assembly and fusion sites of the level 1 construct.

# 2. Materials

## 2.1. Level 0 Plasmids

In the MoClo system, level 0 plasmids contain basic genetic parts (e. g. promoters, RBSs, CDSs, terminators) that can be assembled into TUs, or level 1 constructs via their specific fusion sites (Fig. 1B). Collections of level 0 parts for different organisms, for example, *E. coli* *(9)* or *Vibrio natriegens* *(8)* have been created and can be obtained from Addgene or other sources. Here, we used the Marburg collection *(8)* as a starting point and source for most of the level 0 parts we use. Additionally, to create new, project-specific level 0 parts, we give examples of their construction in 3.1.

## 2.2 Oligonucleotides

Researchers may wish to add their parts of interest. Therefore, to illustrate the design, we show how to create new level 0 plasmids using two examples. Longer parts can be generated by PCR, as shown here for the TetR CDS. Short parts can be created by annealing two oligonucleotides, illustrated here by the promoter pTet. Primer design is illustrated for both cases in Fig. 2. All primer sequences are given in the 5' → 3' direction.

1. Creating level 0 part by PCR:

    (a) Forward primer for level 0 part generated by PCR:

A<u>CGTCTCC</u>*CTCG***NNNN**(n)$_{15-25}$

(b) Reverse primer for level 0 part generated by PCR:

A<u>CGTCTCC</u>*CTCA***NNNN**(n)$_{15-25}$

(c) Example primers used for amplifying the TetR CDS level 0 part (forward/reverse):

A<u>CGTCTCC</u>*CTCG***AATG**(tccagattagataaaagtaaagtg)

A<u>CGTCTCC</u>*CTCA***AAGC**(agacccactttcacatttaagttg)

Where (n)$_{15-25}$ are the nucleotides from the 5' and 3' end of the gene of interest. BsmBI recognition sites are underlined. The overhangs for level 0 part integration into the universal accepting vector (UAV) are italicized. Fusion sites (NNNN) of the Golden Gate level 1 assembly are denoted in bold (*see* **Note 1**).

2. Creating level 0 part by oligo annealing:

(a) Forward oligonucleotide for level 0 part production by annealing:

*CTCG***NNNN**(n)$_{15-max}$

(b) Reverse oligonucleotide for level 0 part production by annealing:

*CTCA***NNNN**(n)$_{15-max}$

(c) Primers for generating the pTet part by oligo annealing (forward/reverse):

*CTCG***GGAG**(ctgttctcagtgatagagattgacatccctatcagtgatagatataatgagcactactagagtact)

*CTCA***AGTA**(ctctagtagtgctcattatatctatcactgatagggatgtcaatctctatcactgagaacagctcc)

Where (n)$_{15-max}$ is the part sequence for annealing into a double stranded product. The overhangs for level 0 integration into UAV are italicized and stay single stranded after annealing of the oligonucleotides. Note that this level 0 part is not cut by BsmBI during level 0 plasmid assembly. The overhangs are already created by annealing. Fusion sites (**NNNN**) of the Golden Gate level 1 assembly are denoted in bold.

3. Primers for linearizing level 1 assembly:

(a) Forward primer for level 0 part generated by PCR:

(N)$_{15-25}$

(b) Reverse primer for level 0 part generated by PCR:

(N)$_{15-25}$

(c) Linearizing primers used in this chapter (forward/reverse):

A1: TGGAACCCAACTCGGAGCTC

A2: ATTGCAGCACTGGGGCCAGATG

Where (N)$_{15-25}$ are sequences from the 5' and 3' ends of the TU. We recommend designing linearizing primers with one binding in the 5' or 3' connector and one in the backbone. Designing two primers binding only in the backbone may result in unspecific amplification of the residual template. For instance, the linear constructs tested in this chapter were obtained by primers binding in the 5' connector and the backbone.

4. Oligonucleotides for annealing the Chi6 DNA:

   (a) Forward oligonucleotide Chi6:
   TCACTTCACTGCTGGTGGCCACTGCTGGTGGCCACTGCTGGTGGCCACTGCTGGTGGCCACTGCTGGTGGCCACTGCTGGTGGCCA

   (b) Reverse oligonucleotide Chi6:
   TGGCCACCAGCAGTGGCCACCAGCAGTGGCCACCAGCAGTGGCCACCAGCAGTGGCCACCAGCAGTGGCCACCAGCAGTGAAGTGA

Forward and reverse Chi6 oligonucleotides are annealed by heating 100 µM of each ssDNA in 1x duplex buffer to 100 °C followed by slow cooling to room temperature (RT). Chi6 dsDNA produced by annealing is stored at -20 °C.

## 2.3 Molecular biology reagents

1. 2x OneTaq DNA polymerase master mix - 2x concentrated mix containing Taq DNA polymerase, dNTPs, $MgCl_2$, buffer components
2. PCR DNA Purification Kit
3. 3 M Sodium acetate solution (pH 5.2)
4. BsaI - Type IIS restriction enzyme
5. BsmBI - Type IIS restriction enzyme
6. T4 DNA Ligase enzyme
7. T4/T7 DNA ligase buffer
8. 1 % Agarose gel
9. 1x TAE (Tris-acetate-EDTA) buffer
10. 10x Duplex buffer: 1M potassium acetate, 300 mM HEPES, pH 7.5

## 2.4 Reagents and solutions for Autolysate preparation

1. 2xYTP Broth: 31 g/L pre-mixed 2xYT (containing 16 g/L tryptone, 10 g/L yeast extract, 5 g/L sodium chloride (NaCl; MW 58.44 g/mol)), 3 g/L Monopotassium phosphate ($KH_2PO_4$; MW 136.09 g/mol), 6.97 g/L Dipotassium phosphate ($K_2HPO_4$; MW 174.18 g/mol). Sterilized by autoclaving.
2. Starter culture: *E. coli* BL21-Gold (DE3) pAD-LyseR (Addgene #99244) - Autolysis *E. coli* strain with ampicillin resistance *(10)*.

3. S30A buffer, pH 7.7: 14 mM Mg-glutamate (L-Glutamic acid hemimagnesium salt tetrahydrate; MW 388.61 g/mol), 60 mM K-glutamate (L-Glutamic acid potassium salt monohydrate; MW 203.23 g/mol), 50 mM Tris (Tris-(hydroxymethyl)-aminomethan; MW 121.14 g/mol).

4. DTT stock solution: 1 M DTT (Dithiothreitol, MW 154.25 g/mol). Store at -20 °C.

5. S30A buffer w/ DTT: S30A buffer + 2 mM DTT.

6. IPTG stock solution: 2 M IPTG (Isopropyl-ß-D-thiogalactopyranosid; MW 238.30 g/mol). Filter to sterilize, and store at -20 °C.

7. D-Glucose stock solution: 2.5 M glucose (D-(+)-Glucose; MW 180.16 g/mol) (use 40-60 °C heating with constant mixing). Filter to sterilize, store at RT.

8. Ampicillin stock solution: 100 mg/ml ampicillin antibiotic (MW 349.40). Filter to sterilize, and store at -20 °C.

## 2.5 Reagents and solutions for Energy Buffer preparation

1. Tris base stock solution: prepare 2 M Tris base solution in water. Filter to sterilize, and store at RT.

2. DTT stock solution: see section 2.4.4.

3. S30A buffer w/ DTT: see section 2.4.5.

4. S30B buffer, pH 8.2: 14 mM Mg-glutamate (L-Glutamic acid hemimagnesium salt tetrahydrate; MW 388.61 g/mol), 60 mM K-glutamate (L-Glutamic acid potassium salt monohydrate; MW 203.23 g/mol), 50 mM Tris (Tris-(hydroxymethyl)-aminomethan; MW 121.14 g/mol). Autoclave to sterilize, store at 4 °C temperature. Before use, add 1 mM DTT.

5. HEPES stock solution: 2 M HEPES (4-(2-Hydroxyethyl)piperazine-1-ethanesulfonic acid; MW 238.30 g/mol). Adjust pH to 8.0 with KOH solution. Filter to sterilize, and store dark and at 4 °C.

*The following reagents need to be prepared freshly for every use, or to be frozen at -80 °C for extended storage:*

6. CoA stock solution: 65 mM CoA (Coenzyme A hydrate; MW 767.53 g/mol).

7. NAD stock solution: 175 mM NAD (β-Nicotinamide adenine dinucleotide hydrate; MW 663.43 g/mol). Adjust pH to 7.5-8 with 2 M Tris.

8. cAMP stock solution: 650 mM cAMP (Adenosine 3′,5′-cyclic monophosphate; MW 329.21 g/mol). Adjust pH to 8.0 with 2 M Tris.

9. Folinic Acid stock solution: 34 mM Folinic acid (Folinic acid calcium salt hydrate; MW 511.50 g/mol).

10. Spermidine stock solution: 1 M spermidine (1,8-Diamino-4-azaoctane; MW 145.25 g/mol) (warm up briefly to 37 °C for better solubility).

11. 3-PGA stock solution: 1.4 M 3-PGA (D-(−)-3-Phosphoglyceric acid disodium salt; MW 230.02 g/mol). Adjust pH to 7.5 with 2 M Tris.

12. Nucleotide Mix stock solution: 156 mM ATP (Adenosine 5′-triphosphate dipotassium salt hydrate; MW 583.36 g/mol), 156 mM GTP (Guanosine 5'-triphosphate disodium salt trihydrate; MW 523.18 g/mol), 94 mM CTP (Cytidine 5'-triphosphate disodium salt trihydrate; MW 483.16 g/mol), 94 mM UTP (Uridine 5'-triphosphate trisodium salt trihydrate; MW 484.14). Adjust pH to 7.5 with 3 M KOH.

13. PEG-8000 stock solution: 40 % PEG-8000 (Poly(ethylene glycol); MW 7000-9000 g/mol). Filter to sterilize.

14. Ammonium glutamate stock solution: prepare 1 M ammonium glutamate (L-Glutamic Acid Ammonium Salt; MW 164.16 g/mol). Filter to sterilize.

15. Oxalic acid stock solution: 1 M oxalic acid (Potassium oxalate monohydrate; MW 184.23 g/mol). Filter to sterilize.

16. Mg-glutamate stock solution: 1 M Mg-glutamate (L-Glutamic acid hemimagnesium salt tetrahydrate; MW 388.61 g/mol). Filter to sterilize.

17. K-glutamate stock solution: 4 M K-glutamate (L-Glutamic acid potassium salt monohydrate; MW 203.23 g/mol). Filter to sterilize.

18. Putrescine stock solution: 1 M putrescine (1,4-Butanediamine dihydrochloride; MW 88.15 g/mol). Filter to sterilize.

## 2.6 Echo® dispensing and plate reader measurement

1. Echo® compatible source plate (e.g. 384PP 2.0, Labcyte)
2. 384-well target plate (e.g. Nunc 384-Shallow Well, black, 264705)
3. Transparent sealing foil (e. g. Ampliseal transparent, Greiner)

# 3 Methods

## 3.1 Construction of new Level 0 plasmids

For level 0 parts that are not included in a preexisting MoClo collection, we advise generating and storing the desired level 0 part in a level 0 plasmid (*see* **Note 2**) (Fig. 2). For primer design see section 2.2 and Fig. 2. We present two methods that can be chosen according to the size of the new level 0 part. For smaller parts (typically under 100 base pairs), it is convenient to directly anneal two oligonucleotides. Longer level 0 parts are generated by PCR amplification of the desired target sequence. If target DNA sequences contain internal BsmBI and/or BsaI restriction sites, these sites need to be removed in advance.

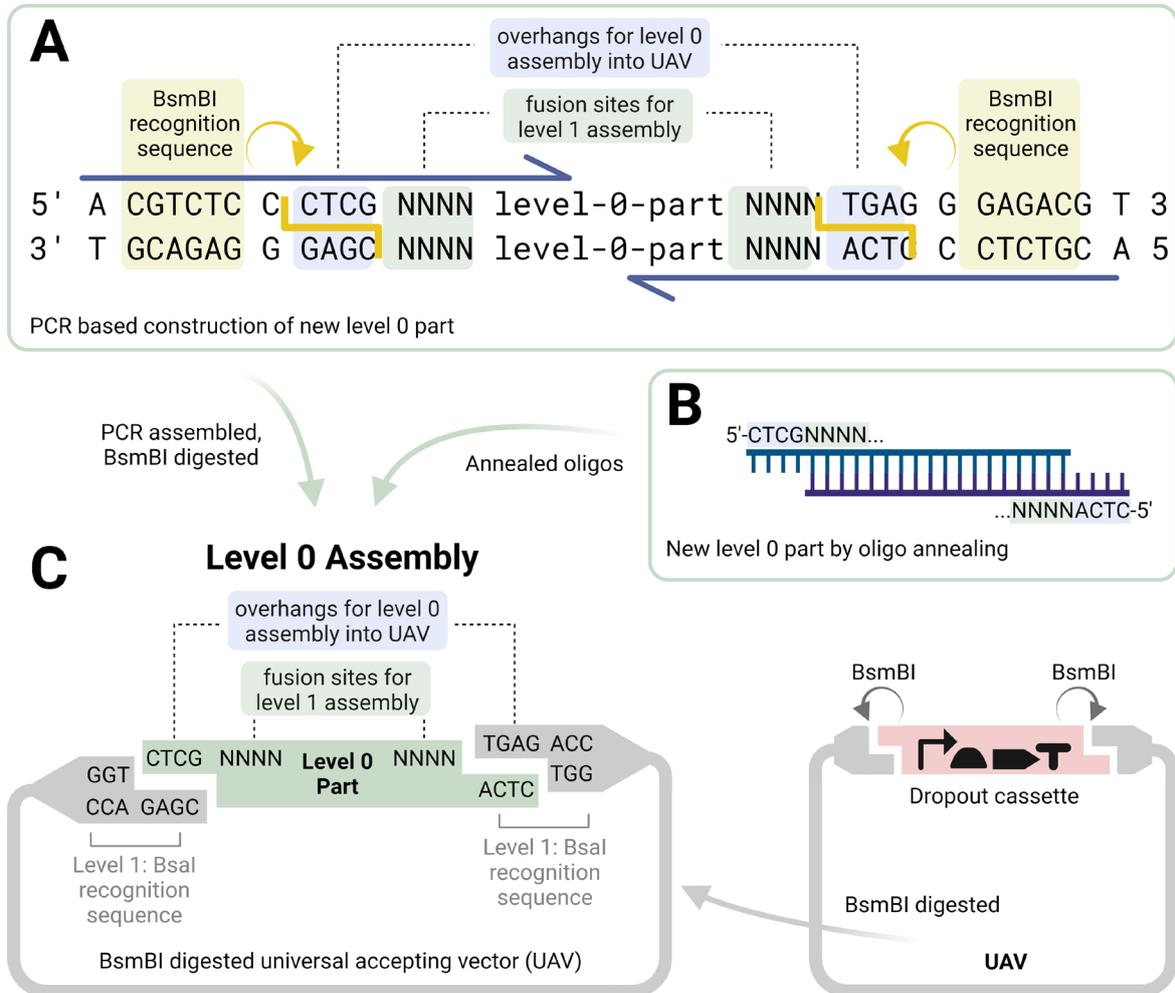

**Fig. 2** Construction of new level 0 parts. A) by PCR. B) by annealing two single-stranded DNA oligonucleotides. C) Parts are assembled into a universal accepting vector (UAV) using BsmBI restriction enzyme.

### 3.1.1 Amplification of a new level 0 part by PCR

In order to amplify a new DNA part of interest, a standard PCR reaction (Fig. 2A) can be run according to the manufacturer's recommendations. 50 µl PCR reaction containing the following components is given as an example:

- 1 µl 10 µM forward primer

- 1 µl 10 µM reverse primer
- 25 µl One*Taq* 2x Master Mix with Standard Buffer
- Template DNA and ddH$_2$O up to 50 µl.

Before further use, PCR products are checked by agarose gel electrophoresis, and then purified using a standard PCR DNA purification kit according to the manufacturer's recommendations.

### 3.1.2. Generating a new level 0 part by annealing oligonucleotides

Short level 0 parts can be rapidly generated by annealing two single stranded DNA oligonucleotides (Fig. 2B). An example of 50 µl reaction mixture for oligonucleotide annealing is given below:

- 1.5 µl Forward oligonucleotide (10 µM)
- 1.5 µl Reverse oligonucleotide (10 µM)
- 5 µl T4 DNA Ligase buffer
- 42 µl ddH$_2$O

Incubate the reaction at 85 °C in a heat block. After incubation, turn off the heat block, and allow the sample to slowly cool down to RT. Annealed oligonucleotides form a fully prepared DNA fragment, which has single-stranded overhangs suitable for Golden Gate assembly.

### 3.1.3. Assembly of new level 0 parts, created by PCR or oligonucleotide annealing

New DNA parts created by PCR or oligonucleotide annealing, can be cloned into the level 0 universal accepting vector (UAV) (Fig. 2C) by using the following Golden Gate reaction mixture:

- 100 ng of DNA insert
- 1 µl UAV vector (25 ng/µl)
- 1 µl T4 DNA Ligase buffer
- 1 µl T4 DNA Ligase
- 0.5 µl BsmBI restriction enzyme
- Up to 10 µl ddH$_2$O

Golden Gate assembly reaction conditions:
50 cycles (37 °C, 2 min; 16 °C, 5 min), followed by 1 cycle at 50 °C, 10 min, and finished with final cycle (80 °C, 10 min , 1 cycle). Finally, hold at 10 °C overnight.

New assemblies are transformed into a suitable cloning strain (such as *E. coli* DH5α) and plated on the appropriate antibiotic. If the accepting vector contains a dropout cassette that results in fluorescent colonies, colonies that likely contain the desired insert can be easily selected by loss of fluorescence (Fig. 2C). New level 0 plasmids are purified by miniprep and sequence verified.

## 3.2 Construction of Level 1 Cell-free Expression Templates

### 3.2.1 Golden Gate reaction for level 1 assembly

Level 1 constructs are assembled by combining several level 0 parts. To build a functional level 1 TU, the user must assemble at least a promoter, a RBS, a CDS, and a terminator. Although it is theoretically possible to use this minimum set of level 0 parts for the level 1 assembly and subsequent linearization, we recommend adding them to a level 1 universal destination vector (UDV) (along with the 5' and 3' connectors) for i) enhancing the assembly efficiency, and ii) enabling storage of the assembly as a plasmid, if desired. Figure 3 gives an overview over the constructs used in this chapter. To demonstrate rapid characterization of expression strengths, we combined five constitutive promoters with two different RBS to generate a small combinatorial library (Fig. 3B).

1. Dilute each level 0 part to 20 fmol/µl. This will help to ensure high assembly efficiency *(6)* and facilitate further handling due to the standardization.

2. Thaw the level 0 parts and the T4 ligase buffer. If your ultimate goal is to create a linear template, you can leave out the backbone (*see* **Note 3**).

3. Each level 0 assembly into level 1 is set up for a final reaction volume of 10 µl and follows this composition:

    - 0.5 µl of each level 0 part at 20 fmol/µl
    - 1 µl 10x T4 ligase buffer
    - 1 µl T4 Ligase
    - 0.5 µl BsaI restriction enzyme
    - ddH2O to 10 µl final reaction volume

    Start by dispensing the level 0 parts that vary through your assemblies (one tube per assembly). Don't dispense the common level 0 parts yet.

4. Prepare the Master Mix assembly according to the final number of reactions. Start by adding ddH2O, then the level 0 parts common to all of your level 1 constructs, the T4 ligase buffer, and the enzymes T4 ligase and BsaI. Mix well and add the master mix to the tubes prepared in step 3.

5. Place your reactions in a thermocycler with the following protocol (see **Note 4**): 15 cycles (37 °C, 1.5 min; 16 °C, 3 min), followed by a final cycle (50 °C, 5 min; 80 °C, 10 min). Finally, hold at 8 °C.

6. After the end of the thermal cycling reaction, dilute the assemblies to 1:10 with ddH2O (see **Note 5**).

7. (*Optional*) Transform your assembly into a suitable cloning strain for plasmid propagation and purification. This is only possible if you added a UDV into the assembly reaction. Storing or testing the assembled plasmids in cells is optional. To generate linear template DNA for cell-free expression, transformation is not necessary.

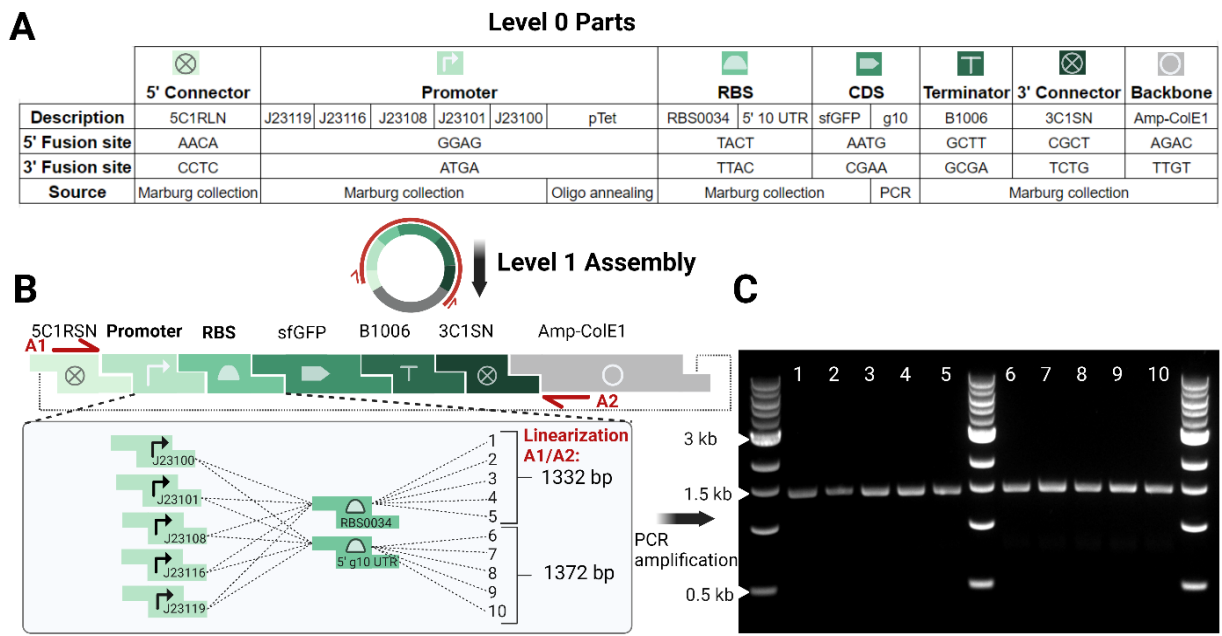

**Fig. 3** Summary of the level 0 and 1 constructs used. A) List of level 0 parts used in this protocol with their origin. Most parts were obtained from the Marburg collection, while the tetR and pTet level 0 parts were prepared by PCR and oligo annealing as described. B) Examples of level 1 assemblies combining five different promoters with two RBS parts, resulting in ten different constructs. C) Amplification of the resulting level 1 constructs with primer A1 and A2 produces DNA products that are suitable linear templates in TXTL reactions.

### 3.2.2 Amplification and purification of level 1 linear templates

1. Set up a 100 μl PCR reaction containing the following components:

    - 2 μl Forward primer A1 (10 μM)

    - 2 μl Reverse primer A2 (10 μM)

    - 50 μl One*Taq* 2x Master Mix with Standard Buffer

    - 2 μl Golden Gate level 1 assembly (diluted 1:10)

- ddH2O to 100 µl final reaction volume

PCR program: Initial denaturation (94 °C, 30 sec), followed by 35 cycles of thermal cycling (94 °C, 30 s; 45-68 °C, 30s; 68 °C, 1 min/kb), then final extension (68 °C, 5 min). Finally, hold at 8 °C.

2. Run the PCR product on a 1 % agarose gel to confirm the correct size. The linearized level 1 construct should be visible as a single band, without side products (Fig. 3C).

3. Proceed with the purification of your linear DNA using a PCR clean-up kit according to the manufacturer's instructions. We recommend using the Invitrogen PureLink PCR Purification Kit (ref. K310001) with the following adjustments for improved binding to the column:

    - 100 µl PCR reaction
    - 100 µl ddH2O
    - 800 µl Binding Buffer B2
    - 10 µl of 3 M Sodium Acetate

To improve DNA yield in the elution step, let the column incubate for up to 1 h at RT after adding the elution buffer, before proceeding to the centrifugation step.

4. Measure the concentration and purity of the PCR products by absorption at 260/280 nm.

5. Send your sample for Sanger sequencing (*see* **Note 6**).

## 3.3 Preparation of TXTL reagents

### 3.3.1 Autolysate preparation

*E. coli* cell lysate for TXTL reactions is prepared from the *E. coli* BL21-Gold (DE3) pAD-LyseR strain. This autolysis strain expresses bacteriophage lambda gene R, causing efficient cell-lysis after freeze-thawing cycle, and produces a highly active lysate for cell-free expression *(10)*.

**Growing and preparing cell biomass**

1. Add ampicillin (Amp, 2.4.8) to the 2xYTP medium (2.4.1) to a final concentration of 100 µg/ml, and inoculate with the autolysis strain glycerol stock (2.4.2). Grow cells overnight at 37 °C, with 300 rpm shaking.

2. Fill two 1 L volume sterilized glass flasks with 400 mL 2xYTP medium each (2.4.1). Add 16 mL of 2.5 M glucose stock solution into each flask (2.4.7) to prepare 2xYTPG medium.

3. Inoculate 400 mL of prepared 2xYTPG medium (from step no. 2) by adding 400 µL Amp stock solution (2.4.8), and 1 mL of autolysis strain overnight culture (from step no. 1). Grow cells at 37 °C, with 300 rpm shaking, until the optical density of the culture (600 nm wavelength, 1 cm path length) $OD_{600}$ = 0.5-0.6.

4. (*Optional*) After reaching $OD_{600}$ = 0.5-0.6, induce cells with 1 mM IPTG (add 200 µL of 2 M IPTG stock solution (2.4.6)). It is only required if you would like to obtain a lysate with active T7 RNA polymerase for expression from T7 promoters.

5. Continue growing the cells at 37 °C, with 300 rpm shaking until $OD_{600}$ = 1.3-1.5 (*see* **Note 7**).

6. Harvest cell biomass: centrifuge cells using a swinging-bucket rotor at 2k x g speed, for 15 min at RT.

7. Remove supernatant. Resuspend biomass from the one batch (400 mL medium) in 45 mL of cold (4−10 °C) S30A buffer (2.4.3) (repeat this step for both grown batches separately).

8. Pre-weigh two empty 50 mL volume falcon tubes and pour two portions of resuspended biomass into them.

9. Centrifuge using the same conditions (see step no. 6).

10. Remove the supernatant carefully, all obvious droplets and liquid leftovers must be cleared away. **Important:** Residual liquid will lead to an overestimation of the biomass amount and result in a lysate with lower expression activity (*see* **Note 8**).

11. Weigh the tubes with centrifuged biomass to determine the amount of biomass per batch. Don't forget to exclude the weight of the empty tubes, measured in step no. 8.

12. Add 2 volumes of cold (4-10 °C) S30A buffer freshly supplemented with 2 mM DTT (2.4.5) according to the weight of cell biomass of each batch (for example, if cell biomass weight is 1235 mg, then 2470 µL of S30A buffer with DTT must be added).

13. Thoroughly resuspend samples by vortexing (to dissolve biomass pellets in the buffer).

14. Freeze cells in buffer at -80 °C. Leave them in the freezer for at least 1 hour (*see* **Note 9).**

**Extracting Autolysate from grown cell biomass**

15. Thaw frozen cells from step no. 14 by leaving them at RT for 10-15 min (*see* **Note 10**).

16. Vortex thawed cells vigorously for at least 5 min. Freeze-thawing and vortexing will lead to cell lysis and produce a very viscous mixture.

17. Incubate biomass at 37 °C, with 300 rpm shaking for 90 min. Shortly vortex tubes after 45 min of incubation.

18. Combine lysed biomass in small centrifugation tubes suitable for high-speed ultracentrifugation. This step may be difficult due to the viscous nature of the mixture (*see* **Note 11**).

19. Centrifuge samples for 45-60 min at 4 °C temperature in an ultracentrifuge with a fixed-angle rotor using 45-50k x g speed (*see* **Note 12**).

20. Very gently move the supernatant to new 1.5 mL tubes (this is your autolysate extract), avoid transferring any pellet or other solid materials (*see* **Note 13**).

21. Centrifuge autolysate extract at RT for 3 min at the maximum speed of a table centrifuge (~21k x g), to remove leftovers of larger particles.

22. Combine the cleared autolysate in one fresh tube and mix well. Then aliquot the autolysate extract into small tubes. Depending on the planned experiments, we recommend aliquot sizes of about 50 µl.

23. Freeze aliquots at -80 °C for storage.

### 3.3.2 Energy buffer preparation

The autolysate extract needs to be supplemented with additional energy buffer components to sustain the reaction. Our TXTL energy buffer recipe is adapted from *(11)*.

**Amino Acid Solution Preparation (6 mM)**

Each amino acid in the stock is usually supplied at 1.5 ml, 168 mM concentration, except for leucine, which is supplied as a 140 mM solution. The final composition of the Amino Acid Solution contains 6 mM of all the amino acids, except leucine (which is at 5 mM concentration).

1. Take all 20 amino acids from -20 °C and thaw them at RT. Vortex until amino acids dissolve (*see* **Note 14**).

2. After dissolving amino acids, put them on ice, **except for Asn, Phe,** and **Cys**: these three amino acids should be kept at RT.

3. Add sterile water to a sterile tube, and place it on ice.

4. To the pre-cooled sterile water, add the necessary amount of each amino acid in the **following order**: Ala, Arg, Asn, Asp, Gln, Glu, Gly, His, Ile, Lys, Met, Phe, Pro, Ser, Thr, Val, Trp, Tyr, Leu, Cys (*see* **Note 15**).

5. Aliquot Amino Acid Solution on ice in sterile tubes. Vortex frequently to avoid unequal distribution of suspension.

6. Store at -80 °C.

**Energy Solution Preparation (14x)**

The final composition of the Energy Solution is 700 mM of HEPES (pH = 8), 21 mM of ATP, 21 mM of GTP, 12.6 mM of CTP, 12.6 mM of UTP, 3.64 mM of CoA, 4.62 mM of NAD, 10.5 mM of cAMP, 0.95 mM of Folinic Acid, 14 mM of Spermidine, 14 mM Putrescine, and 420 mM of 3-PGA. The prepared Energy Solution is at 14x concentration.

7. Thaw on ice all necessary components. Keep solutions on ice for the rest of the time (*see* **Note 16)**.

8. Place a sterile tube on ice, and mix stock solutions in the **following order**: HEPES, water, nucleotide mix, CoA, NAD, cAMP, Folinic acid, spermidine, putrescine, and 3-PGA. Shortly vortex the tube after each added component.

9. Keep prepared Energy Solution on ice, and aliquot into sterile tubes. Vortex the main stock solution frequently to avoid unequal distribution.

10. Freeze aliquots in liquid nitrogen and store at -80 °C.

**Energy Buffer Preparation**

The final composition of the energy buffer is: 16.24 mM Mg-glutamate, 162.4 mM K-glutamate, 3.48 mM amino acids solution, 2.3x energy solution, 2.32 mM DTT, 23.2 mM ammonium glutamate, 23.2 mM oxalic acid, 4.64 % PEG-8000. The prepared energy buffer will be further mixed with autolysate and DNA samples for Cell-free TXTL reactions (see section 3.4), and, therefore, will be additionally diluted 2.3 times.

11. Place sterile tube on ice. To prepare a working solution of energy buffer, mix together: Mg-glutamate, K-glutamate, amino acids solution, energy solution, DTT, ammonium glutamate, oxalic acid, PEG-8000. Shortly vortex solution after the addition of each item (*see* **Note 17**).

12. Keep prepared Energy Buffer on ice, and aliquot into sterile tubes. Vortex the main stock frequently to avoid unequal distribution.

13. Freeze aliquots in liquid nitrogen and store at -80 °C.

## 3.4 Cell-free expression reactions

Cell lysate and energy buffer for TXTL reactions from section 3.3. can be prepared in large batches, aliquoted, and conveniently stored at -80 °C until the time of the experiment. If stored TXTL reagents and a collection of level 0 parts are available, rapid production of linear level 1 DNA templates by Golden Gate assembly and PCR (section 3.2) makes it possible to go from construct design to measurement within one day. In this section, we describe the setup of TXTL reactions for part measurements. Here, we demonstrate combinatorial testing of promoters and ribosomal binding sites (Fig. 4) and characterization of an inverter circuit using the transcriptional repressor TetR (Fig. 5). All linear parts tested are reported through the expression of a superfolder GFP (sfGFP).

The final TXTL reaction mixture for measurements is composed of the following reagents: 40 % v/v of *E. coli* autolysate extract, 10 mM ammonium glutamate; 7 mM Mg-glutamate; 70 mM K-glutamate; 10 mM oxalic acid; 1.5 mM ATP; 1.5 mM GTP; 0.9 mM UTP; 0.9 mM CTP; 0.068 mg/mL folinic acid; 1.5 mM amino acids; 30 mM 3-PGA; 0.33 mM NAD; 0.26 mM CoA; 1 mM putrescine; 1 mM spermidine; 0.75 mM cAMP; 1 mM DTT; 50 mM HEPES, 2 % PEG 8000, 1 µM of Chi6 annealed primers, and varying concentration of linear DNA. The cell-free components are dispensed with an Echo® 525 liquid handler, except for the lysate, which is manually dispensed at the end of the reaction. The use of the acoustic liquid handler is recommended for high-throughput experiments.

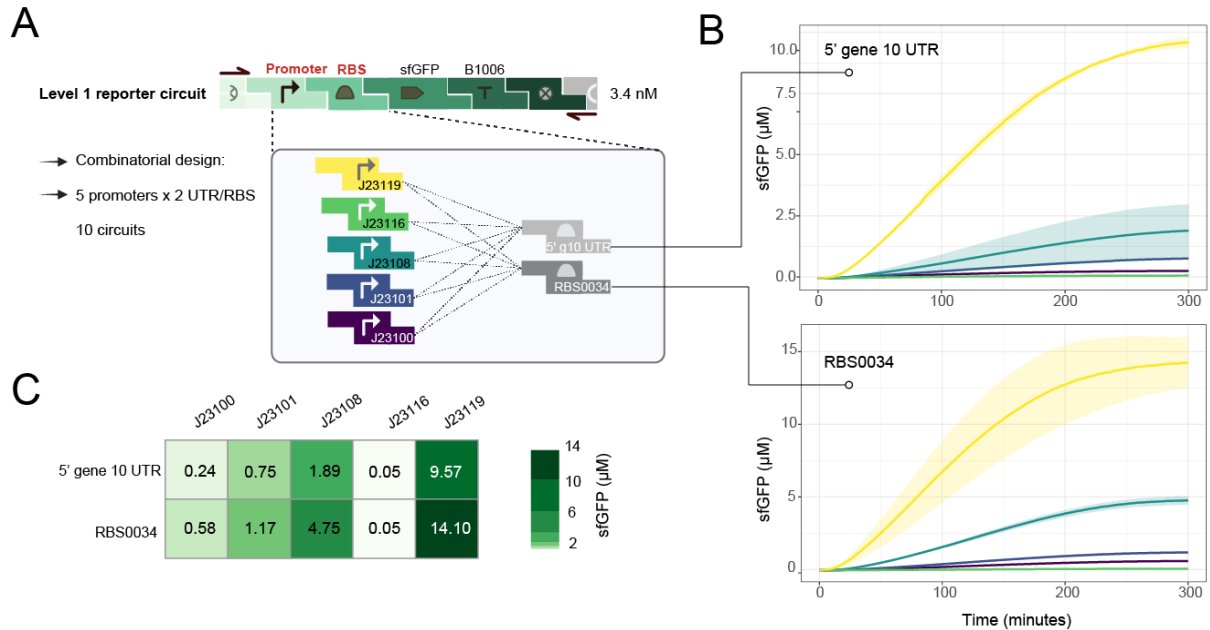

**Fig. 4** Prototyping promoters and RBSs. A. Schematic of parts generated and tested (combinations of multiple promoters and B0034 RBS / T7 gene 10 5' UTR) B. Kinetics of sfGFP production in the cell-free reaction C. Heat map comparing endpoint sfGFP levels.

**Reaction setup and measurement on a plate-reader**

1. Each cell-free reaction is set up for a final volume of 5 µl and is composed of:

    - 2.15 µl energy buffer
    - 2 µl autolysate extract
    - 0.10 µl Chi6 (50 µM)
    - 0 µl to 0.75 µl of concentrated linear DNA
    - ddH2O to 5 µl final reaction volume

    Calculate the number of autolysate extract and energy buffer aliquots needed for your experiment. Take into account reaction replicates and add some spare volume (10%).

    or

    (*Optional*) Use the PyEcho script to generate the Echo® input protocol file according to your experimental design. Refer to the documentation from the GitHub repository for details (https://github.com/HN-lab/PyEcho). Once the experimental design Excel file is

filled, the script will return the amount of reagents needed. Do not include the lysate dispensing. In our experience, autolysate is unsuitable for dispensing with an acoustic liquid handler. Autolysate should be added manually (see step 9).

2. Preheat the plate reader to 29°C.

3. Thaw the cell-free reaction components on ice (DNA linear constructs, Chi6, and the appropriate number of energy buffer and lysate aliquots).

4. Prepare 10x or 5x concentrated DNA solutions containing the linear DNA templates for the TXTL reaction. The final protein production of the TXTL reaction usually saturates at about 10 nM DNA template (final concentration). To determine the effects of promoters or ribosomal binding sites on the expression strength of a reporter gene, we recommend using a near saturating DNA concentration (see Fig. 4). When characterizing the effect of a transcription factor on the expression of a downstream gene in a circuit composed of multiple expression constructs, care should be taken to account for resource competition effects *(12)*. To illustrate the effect of competition for resources resulting from the production of an additional protein, we included a burden reporter with the characterization of the TetR repressor - pTet promoter. The burden reporter consists of a strong non-repressible, constitutive promoter driving the expression of the sfGFP reporter (Fig. 5A). Co-expression of the TetR construct at increasing concentrations decreases the output from the burden reporter, however, less than its target reporter with the pTet promoter (Fig. 5B).

5. (*Only for manual dispensing if Echo® dispenser is not used*) Prepare a mastermix for (n+1) 16.5 µl reactions (each 5 µl cell-free reaction is performed in triplicates and volume loss accounts for 10% of the final volume), consisting of the energy buffer, the autolysate extract, and the Chi6 annealed oligos. First, dispense the concentrated DNA solutions and water in separate reaction tubes. Second, dispense the mastermix in each reaction tube. Third, for each reaction, mix well and pipette 3 times 5 µl in the 384-well plate. Go to step 12.

6. Fill the Echo® source plate with the cell-free components according to your experimental design and the generated protocol. Proceed quickly with the next step.

7. Use an Echo® 525 with the generated input protocol to dispense the concentrated linear DNA parts, the energy buffer, and water into the 384-well target plate (*see* **Note 18**).

8. Seal the plate with a foil to prevent evaporation until the next step (*see* **Note 19**).

9. Remove the cover and manually dispense the lysate into the 384-well plate (*see* **Note 20**).

10. Seal the plate with a transparent foil to prevent evaporation during the plate reader measurement.

11. Shake the plate for a few seconds by stirring manually between two tip boxes.

12. Place the sealed plate into the plate reader and measure every 2 minutes for 5 to 8 hours at the appropriate fluorescence wavelength.

13. Using the data from the plate reader measurement, plot the kinetics of reporter production (Fig. 4B) and quantitate endpoint levels (Fig. 4C, Fig. 5B) to prototype regulatory elements and transcription factors (*see* **Note 21**).

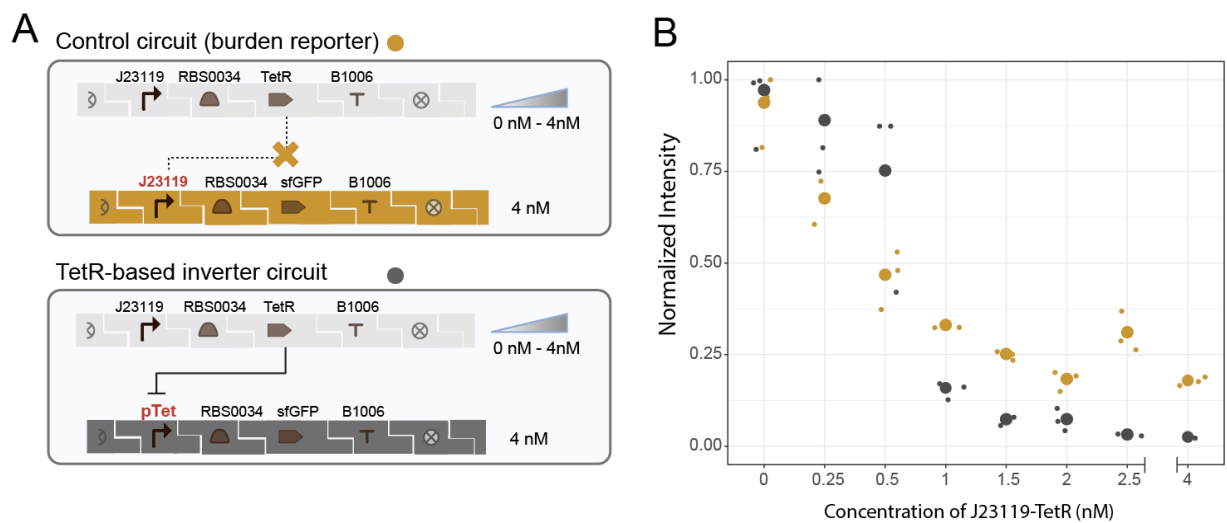

**Fig. 5** Testing a promoter-transcription factor pair. A. Schematic of the burden control circuit and the TetR-based inverter circuit. B. Titration curves of varying the TetR production template DNA concentration in the presence of constant reporter DNA (4 nM). To allow comparison, endpoint fluorescence was normalized to the highest data point for each construct.

# 4 Notes

1. The start and stop codons should be omitted from the primer sequence when designing primers to amplify protein coding sequences for level 0 construction. Start and stop codons are part of the overhang sequence (start codon) or appear during level 1 assembly (stop codon).

2. We advise collecting level 0 parts in plasmids and sequencing the level 0 parts sequences prior to their use. Theoretically, it is also possible to use PCR products as level 0 parts if they contain the correct BsaI recognition and restriction sites. This works well in many cases, but we have seen that for certain parts, it leads to non-specific PCR products during the amplification of the level 1 constructs.

3. We observed a slight decrease in reaction efficiency when assembling without a backbone. We have tested our protocol with up to six level 0s fragments plus backbone (Fig. 2). However, construct assemblies of up to 44 parts in a single reaction have been demonstrated *(13)*. We recommend prior testing when increasing the number of parts in the assembly.

4. No difference was observed in the efficiency of assembly, compared with the longer standard Golden Gate protocol. The improved protocol is about 80 min, whereas the longer standard protocol is about 400 min (see section 3.1.3 for the thermocycling program of the long protocol).

5. It does not matter if the Golden Gate reaction is diluted or not for the PCR reaction. Diluting reaction gives more volume to be used in the future for PCR amplification. The diluted Golden Gate reaction can be stored at -20 °C.

6. Sequencing from the reverse direction through the terminator may sometimes lead to short sequencing reads. This can be solved by changing the placement of the sequencing primers.

7. Measurements of samples with $OD_{600}$ > 1 are not accurate. Therefore, 3-5 fold dilutions of these samples with 2xYTP medium must be prepared to ensure that measurements are performed within the spectrophotometer linear response range.

8. This step is crucial for further lysate preparation. Leftovers of supernatant might dilute the total autolysate batch, affecting transcription−translation activity. You can use aspiration with a vacuum to remove all leftovers of liquids if needed.

9. The freezing step is necessary even if you continue working on the same day. After step no. 14, the preparation procedure can be continued on the next day or even later.

10. You can use a warm water bath (not higher than 37 °C temperature) for 5-10 minutes.

11. After vortexing, partly-lysed cell biomass will be very viscous. Therefore, you can cut the end of a 1 ml pipette tip for easier pipetting to get a wider tip diameter.

12. Fill ¾ of the centrifugation tube volume to prevent the tube from breaking at high speed. For accurate balance, use a second centrifugal tube filled with water. Weigh the balance tube to be exactly the same weight as the corresponding biomass tube.

13. High-speed centrifugation step could also be replaced by longer centrifugation at a lower speed (21-30k x g) *(10)*. However, the higher centrifugation speeds compact the cell-pellet much better and significantly increase the amount of collected supernatant.

14. Shortly incubate at 37 °C for better solubility, if necessary. Cys may not fully dissolve.

15. The order of mixing the amino acids is very important. Vortex the Falcon tube after each addition of amino acid, and keep the solution on ice all the time. Cys can be added as a suspension. After addition, vortex until the solution is relatively clear, incubating at 37 °C if necessary. Cys may not fully dissolve.

16. Each component of the energy solution can be stored at -80 °C as stocks for later use.

17. Energy buffers can be optimized for every batch of autolysate extract by adjusting the final concentrations of Mg-glutamate, K-glutamate, and PEG (**in that order**) to produce cell-free reactions with maximum expression levels. We noticed that DTT calibration does not significantly affect the activity of TXTL reactions. But changes in PEG concentration can have a dramatic influence on activity, especially when used in autolysate-based TXTL reactions, therefore it should be considered carefully.

18. For Echo® dispensing: Acoustic dispensing with the Echo® 525 does not work for solutions containing contamination with long genomic DNA. We suspect that acoustic autolysate dispensing is impossible due to some remaining genomic DNA in the lysate.

19. At this step, you can check the error report of the Echo® Labcyte. If a reagent was not properly dispensed and the reason identified, you can dispense the missing reactants before proceeding to the next step.

20. A dispenser pipette can accelerate the manual dispensing of the autolysate. We advise reducing the duration of this step as much as possible to maximize the cell-free reaction's yield.

21. Alternatively, you can compute sfGFP production rates from the steady-state phase instead of endpoint measurements for benchmarking the performance of constructs. It can help with leaky constructs that accumulate over time.

**Acknowledgments:**

This work was supported by Deutsche Forschungsgemeinschaft (DFG, German Research Foundation) grant NI 2040/1-1. Figures created with BioRender.com.

# References:

1. Silverman AD, Karim AS, and Jewett MC (2020) Cell-free gene expression: an expanded repertoire of applications. Nat Rev Genet 21:151–170

2. Sun ZZ, Yeung E, Hayes CA, et al (2014) Linear DNA for Rapid Prototyping of Synthetic Biological Circuits in an Escherichia coli Based TX-TL Cell-Free System. ACS Synth Biol 3:387–397

3. Marshall R, Maxwell CS, Collins SP, et al (2017) Short DNA containing χ sites enhances DNA stability and gene expression in E. coli cell-free transcription–translation systems. 114:2137–2141

4. Geertz M, Rockel S, and Maerkl SJ (2012) A High-Throughput Microfluidic Method for Generating and Characterizing Transcription Factor Mutant Libraries, In: Weber, W. and Fussenegger, M. (eds.) Synthetic Gene Networks: Methods and Protocols, pp. 107–123 Humana Press, Totowa, NJ

5. Niederholtmeyer H, Stepanova V, and Maerkl SJ (2013) Implementation of cell-free biological networks at steady state. PNAS 110:15985–15990

6. Engler C, Kandzia R, and Marillonnet S (2008) A One Pot, One Step, Precision Cloning Method with High Throughput Capability. PLOS ONE 3:e3647

7. Weber E, Engler C, Gruetzner R, et al (2011) A Modular Cloning System for Standardized Assembly of Multigene Constructs. PLOS ONE 6:e16765

8. Stukenberg D, Hensel T, Hoff J, et al (2021) The Marburg Collection: A Golden Gate DNA Assembly Framework for Synthetic Biology Applications in Vibrio natriegens. ACS Synth Biol 10:1904–1919

9. Moore SJ, Lai H-E, Kelwick RJR, et al (2016) EcoFlex: A Multifunctional MoClo Kit for E. coli Synthetic Biology. ACS Synth Biol 5:1059–1069

10. Didovyk A, Tonooka T, Tsimring L, et al (2017) Rapid and Scalable Preparation of Bacterial Lysates for Cell-Free Gene Expression. ACS Synth Biol 6:2198–2208

11. Sun ZZ, Hayes CA, Shin J, et al (2013) Protocols for implementing an Escherichia coli based TX-TL cell-free expression system for synthetic biology. J Vis Exp e50762

12. Marshall R and Noireaux V (2019) Quantitative modeling of transcription and translation of an all- E. coli cell-free system. Sci Rep 9:11980

13. Werner S, Engler C, Weber E, et al (2012) Fast track assembly of multigene constructs using Golden Gate cloning and the MoClo system. 3:38–43